\documentclass{CFD2017}
\usepackage{CFD2017}
\usepackage{multirow}
\usepackage{bigstrut}
\pdfoutput=1

\title{Adaptive Coarse-Graining for Large-Scale DEM Simulations}
\paperID{CFD 2017-085}
\author{Daniel}{Queteschiner}
\presenting  
\address{CD Laboratory for Multi-Scale Modelling of Multiphase Processes, 4040 Linz, AUSTRIA\\$^2$Department of Particulate Flow Modelling, Johannes Kepler University Linz, 4040 Linz, AUSTRIA\newline
$^3$Linz Institute of Technology (LIT), Johannes Kepler University Linz, 4040 Linz, AUSTRIA}
\email{daniel.queteschiner@jku.at}
\author{Thomas}{Lichtenegger}
\sameaddressas{2,3}
\email{thomas.lichtenegger@jku.at}
\author{Simon}{Schneiderbauer}
\sameaddressas{1}
\email{simon.schneiderbauer@jku.at}
\author{Stefan}{Pirker}
\sameaddressas{2}
\email{stefan.pirker@jku.at}

\begin{document}
\maketitle  
\headers   
\pagenumbering{gobble}

\abstract{
The large time and length scales and, not least, the vast number of particles involved in industrial-scale simulations inflate the computational costs of the Discrete Element Method (DEM) excessively. Coarse grain models can help to lower the computational demands significantly. However, for effects that intrinsically depend on particle size, coarse grain models fail to correctly predict the behaviour of the granular system.

To solve this problem we have developed a new technique based on the efficient combination of fine-scale and coarse grain DEM models. The method is designed to capture the details of the granular system in spatially confined sub-regions while keeping the computational benefits of the coarse grain model where a lower resolution is sufficient. To this end, our method establishes two-way coupling between resolved and coarse grain parts of the system by volumetric passing of boundary conditions. Even more, multiple levels of coarse-graining may be combined to achieve an optimal balance between accuracy and speedup. This approach enables us to reach large time and length scales while retaining specifics of crucial regions. Furthermore, the presented model can be extended to coupled CFD-DEM simulations, where the resolution of the CFD mesh may be changed adaptively as well.
}
\normalfont\normalsize

\printnomenclature[0.7cm]  
\vskip .1em

\section{Introduction}

Since its introduction~\cite{CUND79}, the Discrete Element Method (DEM) has proven to be a viable tool for the analysis of granular flows. Supported by the ever growing computational power, the DEM has found its way into numerous branches of industry such as the minerals and mining industries~\cite{CLEA01}, the transport of consumer goods~\cite{RAJI04}, the pharmaceutical industry~\cite{KETT09}, as well as the iron and steel making industry~\cite{MIO12}.

The major shortcoming of the DEM, however, is its computational cost that increases with the amount of particles involved in the simulation. This hinders the application of the DEM to large-scale systems of industrial size. A coarse grain (CG) model of the DEM has been described~\cite{BIER09,SAKAI09,radl:11:cfd} to improve this situation. Using straightforward scaling rules, a group of particles gets replaced by a representative coarse particle. This effectively reduces the number of particles that need to be processed. The weak point of this approach is that the scaling rules break down when effects depending on the particle size determine the behaviour of the system. Unfortunately, more often than not, industrial facilities operate at multiple scales. 
Hence, for such large-scale simulations, a method is needed, that combines the speedup of the coarse grain model and the resolution of a fine-scale simulation in critical regions. To this end, we propose a concurrent coupling for DEM simulations of different resolution, where one or more fine-scale domains can be embedded in the coarse grain simulation of the overall system.

Indeed, coupling simulations of different resolution or applying models to correct a coarse simulation is not an unusual approach to bridge the scale-gap~\cite{PRAP05,ROJE07,WELL12,SCHN12,SCHN13,SCHN15}.


\section{Model Description}

\subsection{Discrete Element Method}
In the DEM each particle $i = 1, \dots, N$ is advanced in time according to Newton's equations of motion
\begin{eqnarray} \label{eq:Newton}
m_{i} \mathbf{\ddot{x}}_{i} &=& \mathbf{f}_{i}\\
\underline{I_{i}} \bm{\dot{\omega}}_{i} &=& \mathbf{t}_{i}
\end{eqnarray}
The total force $\mathbf{f}_{i}$ acting on a particle includes external forces such as gravity, as well as the normal and tangential contact forces due to binary collisions:
\begin{eqnarray}
\mathbf{f}_{n,ij} &=& k_{n}  \bm{\delta}_{n,ij} - \gamma _{n} \bm{\dot{\delta} }_{n,ij} \label{eq:contactnormal}\\
\mathbf{f}_{t,ij} &=& k_{t} \bm{\delta}_{t,ij} - \gamma _{t} \bm{\dot{\delta} }_{t,ij} \label{eq:contacttangential}
\end{eqnarray}
The tangential overlap is truncated such that
\begin{equation} \label{eq:Coulomb}
f_{t,ij} \leq \mu f_{n,ij}
\end{equation}
where $\mu$ is a Coulomb-like friction coefficient. The expressions for $k_{n,t}$ and $\gamma_{n,t}$ depend on the applied contact model. In this study we used a non-linear damped Hertzian spring-dashpot model~\cite{TSUJI92,ANTY11}. Thereby, the stiffness and damping coefficients read
\begin{eqnarray} \label{eq:DEMcoefficients}
k_{n} &=& \frac{4}{3} E_{\mathit{eff}} \sqrt{R_{\mathit{eff}} \delta _{n,ij}}  \nonumber \\
\gamma _{n} &=& -\beta \sqrt{5 m_{\mathit{eff}} k_{n}} \nonumber \\
k_{t} &=& 8 G_{\mathit{eff}} \sqrt{R_{\mathit{eff}} \delta _{n,ij}}\\
\gamma _{t} &=& -\beta \sqrt{\frac{10}{3} m_{\mathit{eff}} k_{t}} \nonumber \\
\beta &=& \frac{\ln(e)}{\sqrt{\ln^{2}(e) + \pi^{2}}} \nonumber
\end{eqnarray}

\subsection{Coarse Grain Model}
The coarse grain model of the DEM replaces several particles of original size by a single coarse particle and establishes scaling rules based on the assumption of consistent energy densities~\cite{BIER09,radl:11:cfd}. The scaling rules follow from a dimensional analysis of Eqs.~\eqref{eq:contactnormal} and \eqref{eq:contacttangential} and are applicable to the contact model used in this work~\cite{NASA15}. In detail, the particle density, the coefficient of restitution, the Young's modulus and the coefficient of friction need to be kept constant. The particle radius is scaled with the constant coarse grain ratio $\alpha$.  The stiffness coefficients $k_{n,t}$ scale with $\alpha$ and the damping coefficients $\gamma_{n,t}$ scale with $\alpha^{2}$.

\subsection{Multi-Level Coarse Grain Model}
To combine the advantages of the fine-scale and coarse grain DEM models, we embed one or multiple fine-scale subdomains in the coarse grain simulation using equivalent external forces and the same geometries in any part of the system. This can be done recursively to nest multiple coarse grain levels.

At the boundary surface of the fine-scale region we measure the mass flow rate, the particle velocity and size distribution of the coarse grain particles. To this end, we divide the surface into tetragonal cells that are about three to ten coarse grain diameters in size.

We consider ensembles of particles instead of tracking each individual grain to retain the local size distribution while avoiding the introduction of artificial clusters of equal particles due to a one-to-one replacement of particles.

Over the course of a coupling interval the particle velocity per cell is Favre averaged. We use the data thus obtained to insert the corresponding fine-scale particles cell by cell by means of a simple sequential inhibition (SSI) algorithm~\cite{DIGG76}.

Furthermore, we introduce a boundary layer inside the fine-scale subdomain to establish proper boundary conditions and ensure a smooth transition between the differently resolved representations. Analogous to the boundary surface, this region is subdivided into a single layer of hexahedral cells, which are used to obtain Eulerian properties of the material such as the macroscopic stress \cite{CHIALVO2012},
\begin{equation} \label{eq:granularstress}
\sigma = \frac{1}{V} \sum_{i}\left[\sum_{j\neq i} \frac{1}{2} \mathbf{r}_{ij}\mathbf{f}_{ij} + m_{i}(\mathbf{v}'_{i})(\mathbf{v}'_{i})\right]
\end{equation}
the volume fraction and the average as well as the maximum particle velocities per cell.

To transfer the granular stress from the coarse grain simulation to the fine-scale simulation, a discrete proportional-integral (PI) controller is used with the fine-scale normal stress components as process variable and the corresponding coarse grain properties as setpoint. This results in a correcting force that is applied to the fine-scale particles in the transition layer. We limit the force such that the resulting particle velocity will not exceed the maximum velocity in the master coarse grain simulation. Furthermore, we let the force induced by the mismatch of the normal stress components fade out after $\nicefrac{3}{2}$ of a fine-scale particle diameter, which is sufficient to account for the missing particles at the boundaries.

We could now handle the transition from the fine-scale subdomain to the master coarse-grain simulation in a similar way, thus establishing a symmetric coupling scheme. 
However, this introduces additional particle insertion events, which are typically a costly operation in DEM simulations as they invalidate the current neighbour list. Furthermore, the determination of appropriate insertion locations for coarse grain particles may become a non-trivial task. To ensure the strict conservation of mass, this may involve a complicated gathering step to accumulate fine-scale particles crossing the boundary surface. 
In addition, another transition layer and controller forces may decrease the stability of the system.

To avoid these potential problems, we instead preserve the coarse grain particles and apply correcting forces, if necessary, to ensure an accurate overall behaviour of the coarse grain system. 

These corrections are realized analogous to the establishing of boundary conditions in the transition layer. We introduce a hexahedral grid inside the fine-scale region and thus obtain volume-averaged particle properties. These are fed into a controller to be transferred to the coarse grain particles.
Typically, the property that needs to be corrected is either velocity or mass flow rate, where it may be necessary to sacrifice one over the other to achieve a correct overall behaviour of the coarse grain system outside the fine-scale domain. In case of correcting the mass flow rate, we typically need to increase the velocity of the coarse grain particles due to a reduced volume fraction. Hence, we simply multiply the fine-scale velocity by the volume-fraction mismatch. A simple proportional controller is sufficient for velocity and mass flow adjustments. The advantage over directly setting the velocity is a smoother transition across cells.

Despite the corrections applied to the coarse grain system, at any given location the evaluation of data should be performed using the highest resolution available.


\InsFig{figure1}{Particles filled into the silo (only coarse grain particles shown). The grid indicated at the bottom is used to establish the coupling between the different coarse grain levels.}{silofilled}

\section{Results}
To test the behavior and performance of our multi-level coarse grain implementation, the filling and discharge of a silo was studied. This test case has previously been used to illustrate the performance of an MPI/OpenMP hybrid parallelization of the LIGGGHTS open source DEM code~\citep{BERGER15}.
Considering the Beverloo equation~\citep{BEVER61}
\begin{equation} \label{eq:Beverloo}
\dot{m} = C\rho \sqrt{g}(D_{o} - \kappa d)^{5/2}
\end{equation}
which predicts the discharge rate of monodisperse granular material through a circular orifice, a dependence on the particle size is clearly evident. Thus, we can expect a different behaviour of the coarse grain and the fine-scale simulation.
To testify this prediction, we compare a reference simulation with particles of original size to a conventional coarse grain simulation and a simulation using our model with two levels of resolution.
The reference simulation consists of 187~504 particles with a diameter of 2.8~mm. The coarse grain simulation uses a coarse grain ratio of $\alpha=2$, i.e., 23~438 particles with a diameter of 5.6~mm. Finally, the multi-level coarse grain simulation is constituted of 23~438 particles scaled with $\alpha=2$, and about 44~000 particles of original size in the lower quarter of the silo.
The simulation parameters are given in Table~\ref{tab:parameters}.

The particles are poured into a silo of 40~cm height. The top half of the silo is a cylinder with a diameter of 27~cm, while the lower conical half narrows to a 4~cm diameter. After an incipient filling and settling phase of 0.7~s, the orifice at the bottom is opened, letting the particles flow out for 1.0~s. Figure~\ref{fig:silofilled} shows a cross-section of the silo after the initial phase.

The grid illustrated at the bottom of the silo in Fig.~\ref{fig:silofilled} indicates the fine-scale subdomain of the multi-level setup and is used to obtain volume-averaged quantities for the coupling procedure. The grid is made up of 172 cubic cells with an edge length of 2~cm. The top layer consisting of 60 cells is used for the transition from the coarse-scale to the fine-scale representation of the particles. The cells below are used for mass flow corrections of the coarse-scale simulation. The discharge rate is measured 1~cm below the orifice with a sampling rate of 100~Hz.

All simulations were performed on an Intel Core i5-4570 CPU using a  $2 \times 2 \times 1$ partitioning of the simulation domain.

\begin{table}[htbp]
  \centering
  \caption{Simulation parameters of silo example.}
  \label{tab:parameters}
    \begin{tabular}{|ll|}
    \hline
    \bigstrut[t] Young's modulus  & $2.5 \times 10^{7}$ N/m$^{2}$ \\
    Poisson's ratio  & 0.25 \\
    Coefficient of restitution  & 0.5 \\
    Coefficient of friction (particle-particle)  & 0.2 \\
    Coefficient of friction (particle-wall)  & 0.175 \\
    Particle density  & 1000~kg/m$^{3}$ \\
    Particle diameter  & 2.8~mm \\
    Time step  & $10^{-6}$~s \\
    Duration  & $1.7 \times 10^{6}$ steps \\
    \hline
    \end{tabular}
\end{table}


\subsection*{Silo Filling}

\begin{figure}[!b]
    \includegraphics[scale=1]{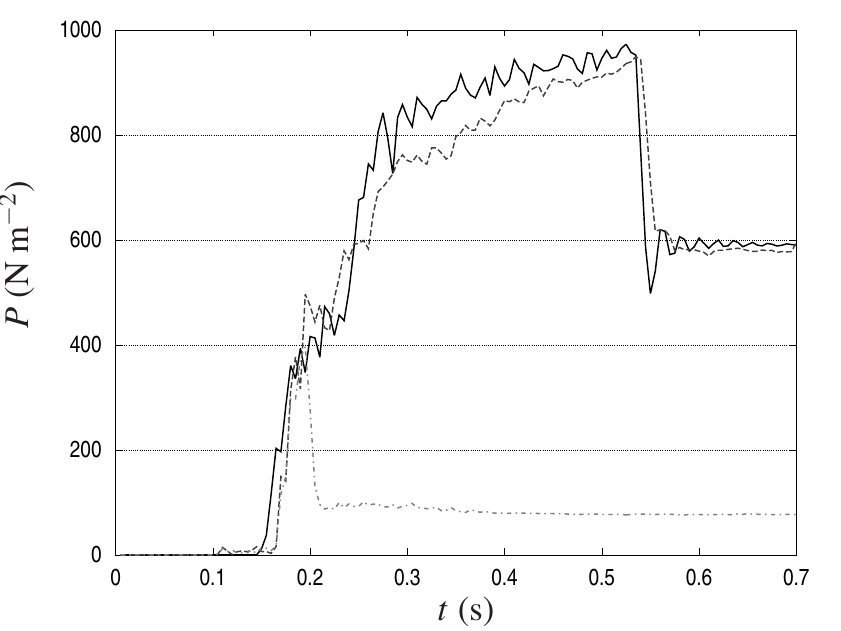}
    \caption{Average granular pressure in N~m$^{-2}$ in the four central cells of the transition layer as a function of time. --- reference simulation, --~-- fine-scale subdomain with pressure correction and $-\cdot-$ without pressure correction.}
\label{fig:pressure_ref_2lvl}
\end{figure}
Concentrating on the filling part of the simulation allows us to
focus on the transition from the coarse-scale representation to the fine-scale description of the system. In the multi-level coarse grain variant of this test case, the original size particles at the top of their subdomain are lacking the pressure exerted from the particles further above. The stress-based PI controller of our model is to correct this deficiency.

Figure~\ref{fig:pressure_ref_2lvl} shows the average granular pressure in the four central cells of the transition region as a function of time for the reference simulation and the fine-scale region of the two-level coarse grain simulation with and without corrections.

To fill up the silo, particles are inserted from $t = 0$~s to $t = 0.47$~s at the top of the silo with an initial velocity of $v_{z} = -3$~m/s. From the diagram in Fig.~\ref{fig:pressure_ref_2lvl} we find a change in pressure around $t = 0.15$~s. At this point, the heap of particles reaches the lower boundary of the transition region.
Between $t = 0.15$~s and $t = 0.2$~s the transition region is filled up. For the uncorrected fine-scale subdomain, the pressure drops at this point and levels off at about 77~N/m$^{2}$. The reference simulation, however, shows a further increase of the pressure due to additional particles falling on top of the fill. The pressure drop at $t = 0.55$~s marks the start of the settling phase where all particles come to rest. This is accompanied by a pressure relaxation.

By applying the granular stress from the master coarse-scale simulation via the PI controller with an update every 15 time steps, this behaviour can be reproduced in good agreement in the embedded fine-scale simulation.


\subsection*{Silo Discharge}

\begin{figure}[!b]
    \includegraphics[scale=1]{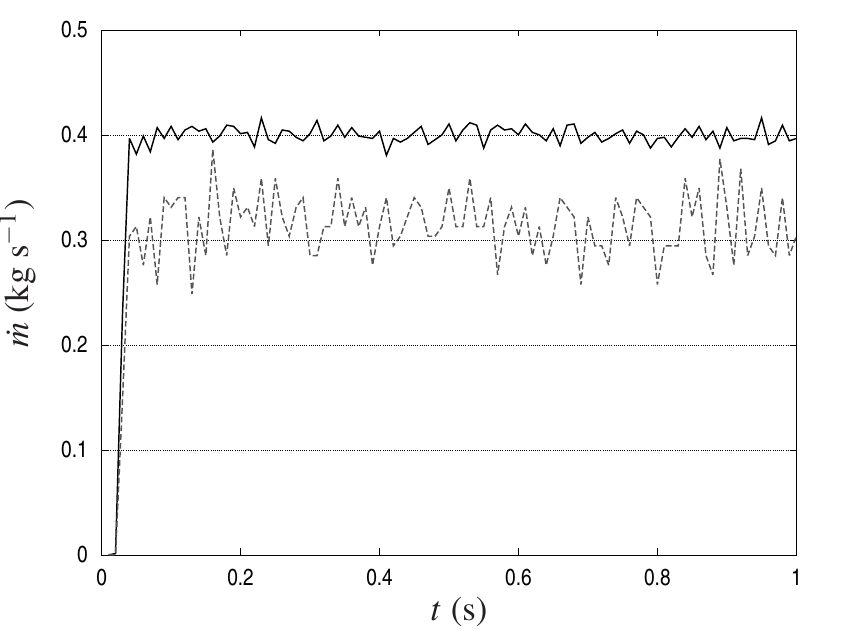}
    \caption{The rate of discharge in kg~s$^{-1}$ of 2.8~mm particles as a function of time. --- reference simulation and --~-- coarse grain simulation ($\alpha=2$)}
\label{fig:massflowrate_ref_cg2}
\end{figure}

\begin{figure}[!b]
    \includegraphics[scale=1]{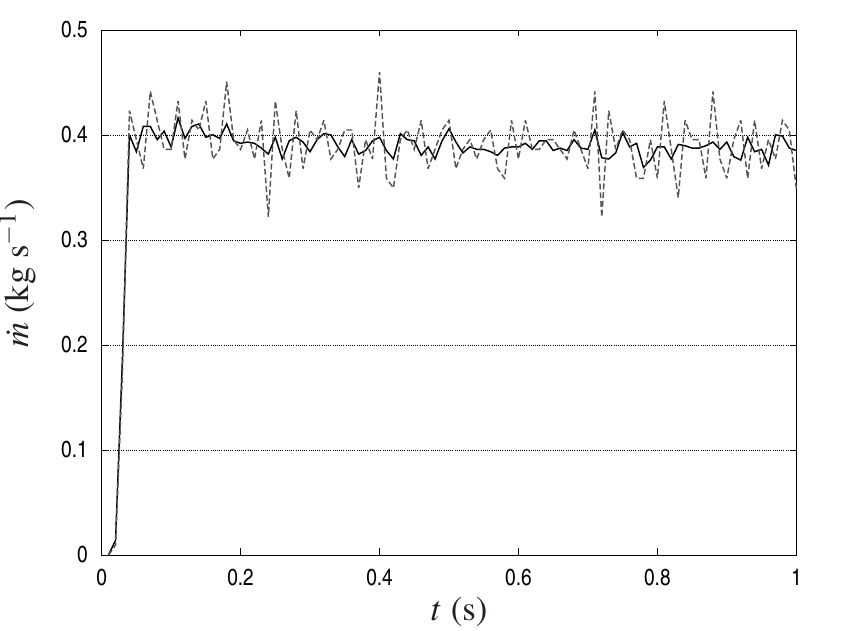}
    \caption{The rate of discharge in kg~s$^{-1}$ of 2.8~mm particles as a function of time. Two size levels: --- original size particles and --~-- coarse grain particles ($\alpha=2$)}
\label{fig:massflowrate_2lvl}
\end{figure}

The discharge phase of the simulation lends itself to study the  correction of the coarse grain simulation using properties of the particles in the fine-scale subregion. Based on Eq.~\eqref{eq:Beverloo}, we can assume that the conventional coarse grain model will fail to correctly predict the discharge rate from the silo.

Indeed, the computed mass flow rates of the reference and coarse grain simulations depicted in Fig.~\ref{fig:massflowrate_ref_cg2} confirm that the setup containing the coarse particles exhibits a substantially lower discharge rate. While the reference simulation yields an average of 0.4~kg/s, the coarse grain equivalent underpredicts the mass flow rate with 0.315~kg/s by more than 20\% (cf. Table~\ref{tab:massflowrate}). In addition, we observe larger fluctuations in the discharge rate of the coarse grain simulation. These are mainly due to the discrete nature of mass in the granular material. The detection of a single particle in the coarse grain simulation corresponds to $\alpha^{3}$ particles in the fine-scale description. On the other hand, the time span between detection events of coarse grains is larger due to their increased size and thus an increased distance between individual particles in flow direction.

Also, due to kinematic constraints, the volume fraction in the coarse grain simulation is lower near the outlet. Thus, for the coupling method in our model it is insufficient to solely correct the particle velocity. It is essential to take the volume fraction mismatch into account as an additional parameter. In fact, reaching the same velocity and mass flow rate in the coarse grain simulation and the fine-scale simulation at the same time is mutually exclusive in this region. In this regard, it should be noted that once the coarse-scale particles have passed the orifice, a pure velocity coupling is to be applied to ensure a correct flow velocity in any adjacent coarse-scale regions.

Applying this procedure, an average discharge rate of 0.39~kg/s was computed in our two-level simulation (cf. Table~\ref{tab:massflowrate}). This means no more than 2.5\% deviation from the reference value. Figure~\ref{fig:massflowrate_2lvl} depicts the discharge rate as a function of time and shows that the coarse grain part of the coupled simulation follows the fine-scale part closely. Also, the fluctuations of the discharge rate in the coupled simulation are comparable to those in the corresponding reference and coarse grain simulations. Hence, we can conclude that our controlling scheme does not add any significant noise to the flow characteristics.

     \begin{table}[htbp]
\centering
\caption{Computed averaged discharge rates $\langle \dot{m}\rangle$ with corresponding standard deviation $\sigma \left( \dot{m}\right) = \sqrt{\langle \left(\dot{m} - \langle \dot{m}\rangle\right)^{2}\rangle}$ in kg s$^{-1}$. The speedup of the simulation runtime is given relative to the reference simulation.}
\label{tab:massflowrate}
\begin{tabular}{|l*{2}c{r}|}
\hline
\bigstrut & $\langle \dot{m}\rangle$ & $\sigma \left( \dot{m}\right)$ & speedup \\
\hline
\bigstrut[t] Reference Simulation & 0.400 & 0.007 & 1.0$\times$    \\ 
CG Model ($\alpha=2$)        & 0.315 & 0.028 & 10.1$\times$ \\
MLCG Model ($\alpha=1$)      & 0.390 & 0.009 & \multirow{2}{*}{2.4$\times$ }  \\
MLCG Model ($\alpha=2$) & 0.391 & 0.026 &  \\
\hline
\end{tabular}
\end{table}

\subsection*{Simulation Runtimes}

A list of the relative runtimes of the reference, conventional coarse grain and multi-level coarse grain simulations is given in Table~\ref{tab:massflowrate}. We note that the $10.1\times$ speedup of the conventional coarse grain simulation is slightly higher than one may expect from the ratio of the number of coarse grain particles to the number of fine-scale particles. Although it is difficult to determine the exact source for this additional speedup, a somewhat reduced number of average neighbour particles and fewer neighbour list rebuilds in the coarse grain simulation are assumed to contribute to the effect.

In the multi-level coarse grain simulation, the total number of particles after filling is $2.78\times$ lower than in the reference simulation. However, this ratio gets worse during discharge, as the net amount of particles in the lower quarter of the silo does not change significantly. Furthermore, the insertion of particles into the fine-scale subdomain, which occurs at regular intervals, triggers additional neighbour list rebuilds adding to the runtime. Also, the calculation of the cell-averaged particle properties adds a minor overhead. Hence, the measured speedup of $2.4\times$ comes up to expectations, especially when we consider the low coarse grain ratio $\alpha$ and the - for demonstration purpose - exaggerated fine-scale region in the presented test case.

A more realistic scenario may be imagined by reducing the particle size by a factor of 32 and conversely increasing the amount of particles by 32$^3$, resulting in about 6.14 billion particles in the full system. Assume we establish a recursive coupling of five coarse grain levels $l = 1,\ldots,5$ with $\alpha_l = 2^{l-1}$. Furthermore, let the subdomains be defined such that the volume filled with particles is quartered compared to the next coarser level. This means that in each level $l$ we end up with about $6.14\times 10^9 \times \alpha_l^{-3} \times 4^{l-5}$ particles. This amounts to approximately 46.5 million particles in total and we may estimate a speedup of more than $100\times$.

The runtime of the simulation can be further improved when taking into account the dependency of the time step on the particle size. In the DEM, the time step needs to be chosen such that the overlap of particles during contact can be resolved. This implies that, in accordance with the particle size, the time step may be scaled with $\alpha$. 
Ultimately, the speedup depends on the size of the region required to resolve the critical area in the system, as well as the desired level of accuracy.


\section{Conclusion}

We described a new technique to concurrently simulate granular flows at different coarse grain levels, where spatially confined subdomains of finer scale are embedded into coarser representations of the system. We presented data to confirm the proper establishing of boundary conditions for the fine-scale region. This was achieved by applying stress-based controller forces within a predefined transition region. Furthermore, we demonstrated that the more precise data of the fine-scale subdomain can be used to amend the overall behaviour of the coarse-scale simulation. We have validated the method by comparing the computed Eulerian properties of the multi-level coarse grain model with the corresponding properties of the fully resolved reference system.

The computational speedup in the presented test case was nearly proportional to the number of particles saved. This means that our method introduces only a minor overhead compared to the overall computational costs per particle. As the amount of particles is generally the major limiting factor, our method performs best for systems that require full resolution only in small regions of the simulation domain and allow for large coarse grain ratios in the rest of the system.

The presented method can be easily extended to improve the performance of coupled CFD-DEM simulations, where the DEM component typically takes up the major part of the computational resources.
The different coarse grain representations of the granular material can be treated separately on the CFD side using appropriately scaled drag laws. The DEM part can then merge the different levels as demonstrated in this study. Furthermore, the resolution of the CFD mesh can be chosen according to the DEM coarse grain level in the corresponding region.

\section{Acknowledgements}

This work was funded by the Christian-Doppler Research Association, the Austrian Federal Ministry of Economy, Family and Youth, and the Austrian National Foundation for Research, Technology and Development. Furthermore, the authors want to thank the K1-MET center for metallurgical research in Austria, which is partly funded by the Austrian government (www.ffg.at), for its financial contribution.

\bibliographystyle{CFD2017}
\bibliography{cfd2017_queteschiner}   

\begin{thebibliography}{21}
\newcommand{\enquote}[1]{``#1''}
\providecommand{\natexlab}[1]{#1}
\providecommand{\url}[1]{\texttt{#1}}
\providecommand{\urlprefix}{URL }
\providecommand{\eprint}[2][]{\url{#2}}

\bibitem[{Antypov and Elliott(2011)}]{ANTY11}
ANTYPOV, D. and ELLIOTT, J.A. (2011).
\newblock \enquote{{On an analytical solution for the damped Hertzian spring}}.
\newblock \emph{EPL}, \textbf{94(5)}, 50004.

\bibitem[{Berger \emph{et~al.}(2015)}]{BERGER15}
BERGER, R., KLOSS, C., KOHLMEYER, A. and PIRKER, S. (2015).
\newblock \enquote{{Hybrid parallelization of the LIGGGHTS open-source DEM
  code}}.
\newblock \emph{Powder Technol.}, \textbf{278}, 234--247.

\bibitem[{Beverloo \emph{et~al.}(1961)}]{BEVER61}
BEVERLOO, W., LENIGER, H. and VAN~DE VELDE, J. (1961).
\newblock \enquote{The flow of granular solids through orifices}.
\newblock \emph{Chem. Eng. Sci.}, \textbf{15(3-4)}, 260--269.

\bibitem[{Bierwisch \emph{et~al.}(2009)}]{BIER09}
BIERWISCH, C., KRAFT, T., RIEDEL, H. and MOSELER, M. (2009).
\newblock \enquote{Three-dimensional discrete element models for the granular
  statics and dynamics of powders in cavity filling}.
\newblock \emph{J. Mech. Phys. Solids}, \textbf{57(1)}, 10--31.

\bibitem[{Chialvo \emph{et~al.}(2012)}]{CHIALVO2012}
CHIALVO, S., SUN, J. and SUNDARESAN, S. (2012).
\newblock \enquote{Bridging the rheology of granular flows in three regimes}.
\newblock \emph{Phys. Rev. E}, \textbf{85}, 021305.

\bibitem[{Cleary(2001)}]{CLEA01}
CLEARY, P. (2001).
\newblock \enquote{{Modelling comminution devices using DEM}}.
\newblock \emph{Int. J. Numer. Anal. Meth. Geomech.}, \textbf{25(1)}, 83--105.

\bibitem[{Cundall and Strack(1979)}]{CUND79}
CUNDALL, P.A. and STRACK, O.D.L. (1979).
\newblock \enquote{A discrete numerical model for granular assemblies}.
\newblock \emph{G\'{e}otechnique}, \textbf{29(1)}, 47--65.

\bibitem[{Diggle \emph{et~al.}(1976)}]{DIGG76}
DIGGLE, P.J., BESAG, J. and GLEAVES, J.T. (1976).
\newblock \enquote{{Statistical Analysis of Spatial Point Patterns by Means of
  Distance Methods}}.
\newblock \emph{Biometrics}, \textbf{32(3)}, 659--667.

\bibitem[{Ketterhagen \emph{et~al.}(2009)}]{KETT09}
KETTERHAGEN, W.R., AM~ENDE, M.T. and HANCOCK, B.C. (2009).
\newblock \enquote{{Process Modeling in the Pharmaceutical Industry using the
  Discrete Element Method}}.
\newblock \emph{J. Pharm. Sci.}, \textbf{98(2)}, 442--470.

\bibitem[{Mio \emph{et~al.}(2012)}]{MIO12}
MIO, H., KADOWAKI, M., MATSUZAKI, S. and KUNITOMO, K. (2012).
\newblock \enquote{{Development of particle flow simulator in charging process
  of blast furnace by discrete element method}}.
\newblock \emph{Miner. Eng.}, \textbf{33}, 27--33.

\bibitem[{Nasato \emph{et~al.}(2015)}]{NASA15}
NASATO, D.S., GONIVA, C., PIRKER, S. and KLOSS, C. (2015).
\newblock \enquote{{Coarse Graining for Large-scale DEM Simulations of Particle
  Flow - An Investigation on Contact and Cohesion Models}}.
\newblock \emph{Procedia Eng.}, \textbf{102}, 1484--1490.

\bibitem[{Praprotnik \emph{et~al.}(2005)}]{PRAP05}
PRAPROTNIK, M., DELLE~SITE, L. and KREMER, K. (2005).
\newblock \enquote{{Adaptive resolution molecular-dynamics simulation: Changing
  the degrees of freedom on the fly}}.
\newblock \emph{J. Chem. Phys.}, \textbf{123(22)}, 224106.

\bibitem[{Radl \emph{et~al.}(2011)}]{radl:11:cfd}
RADL, S., RADEKE, C., KHINAST, J. and SUNDARESAN, S. (2011).
\newblock \enquote{{Parcel-Based Approach for the Simulation of Gas-Particle
  Flows}}.
\newblock J.E.{\O}. Olsen and S.T. Johansen (eds.), \emph{{Proceedings of the
  8th International Conference on CFD in Oil \& Gas, Metallurgical and Process
  Industries}}, 124/1--124/10. Flow Technology.

\bibitem[{Raji and Favier(2004)}]{RAJI04}
RAJI, A. and FAVIER, J. (2004).
\newblock \enquote{{Model for the deformation in agricultural and food
  particulate materials under bulk compressive loading using discrete element
  method. I: Theory, model development and validation}}.
\newblock \emph{J. Food Eng.}, \textbf{64(3)}, 359--371.

\bibitem[{Rojek and O{\~n}ate(2007)}]{ROJE07}
ROJEK, J. and O{\~N}ATE, E. (2007).
\newblock \enquote{Multiscale analysis using a coupled discrete/finite element
  model}.
\newblock \emph{Interact. Multiscale Mech.}, \textbf{1(1)}, 1--31.

\bibitem[{Sakai and Koshizuka(2009)}]{SAKAI09}
SAKAI, M. and KOSHIZUKA, S. (2009).
\newblock \enquote{Large-scale discrete element modeling in pneumatic
  conveying}.
\newblock \emph{Chem. Eng. Sci.}, \textbf{64(3)}, 533--539.

\bibitem[{Schneiderbauer \emph{et~al.}(2012)}]{SCHN12}
SCHNEIDERBAUER, S., AIGNER, A. and PIRKER, S. (2012).
\newblock \enquote{{A comprehensive frictional-kinetic model for gas-particle
  flows: Analysis of fluidized and moving bed regimes}}.
\newblock \emph{Chem. Eng. Sci.}, \textbf{80}, 279--292.

\bibitem[{Schneiderbauer \emph{et~al.}(2013)}]{SCHN13}
SCHNEIDERBAUER, S., PUTTINGER, S. and PIRKER, S. (2013).
\newblock \enquote{Comparative analysis of subgrid drag modifications for dense
  gas-particle flows in bubbling fluidized beds}.
\newblock \emph{AIChE J.}, \textbf{59(11)}, 4077--4099.

\bibitem[{Schneiderbauer \emph{et~al.}(2015)}]{SCHN15}
SCHNEIDERBAUER, S., PUTTINGER, S., PIRKER, S., AGUAYO, P. and KANELLOPOULOS, V.
  (2015).
\newblock \enquote{{CFD modeling and simulation of industrial scale olefin
  polymerization fluidized bed reactors}}.
\newblock \emph{Chem. Eng. J.}, \textbf{264}, 99--112.

\bibitem[{Tsuji \emph{et~al.}(1992)}]{TSUJI92}
TSUJI, Y., TANAKA, T. and ISHIDA, T. (1992).
\newblock \enquote{Lagrangian numerical simulation of plug flow of cohesionless
  particles in a horizontal pipe}.
\newblock \emph{Powder Technol.}, \textbf{71(3)}, 239--250.

\bibitem[{Wellmann and Wriggers(2012)}]{WELL12}
WELLMANN, C. and WRIGGERS, P. (2012).
\newblock \enquote{A two-scale model of granular materials}.
\newblock \emph{Comput. Meth. Appl. M.}, \textbf{205-208}, 46--58.

\end{thebibliography}

\end{document}